\documentclass[aps,pre,reprint,superscriptaddress,nofootinbib]{revtex4-1}
\usepackage{graphicx}
\usepackage{amsmath}
\usepackage{amsfonts}
\usepackage{hyperref}
\begin{document}

\renewcommand\vec[1]{\mathbf{#1}}
\newcommand{\Det}[1]{\ensuremath{\textrm{det} \left( #1 \right)}}
\newcommand{\brm}[1]{\bm{{\rm #1}}}
\newcommand{\Cv}{\bm{{\rm C}}}
\newcommand{\Qv}{\bm{{\rm Q}}}
\newcommand{\qv}{\bm{{\rm q}}}
\newcommand{\fv}{\bm{{\rm f}}}
\newcommand{\ev}{\bm{{\rm e}}}
\newcommand{\tv}{\bm{{\rm t}}}
\newcommand{\uv}{\bm{{\rm u}}}
\newcommand{\kv}{\bm{{\rm k}}}
\newcommand{\Rv}{\bm{{\rm R}}}
\newcommand{\Dv}{\bm{{\rm D}}}
\newcommand{\av}{\bm{{\rm a}}}
\newcommand{\bv}{\bm{{\rm b}}}
\newcommand{\Gv}{\bm{{\rm G}}}

\title{Topological boundary modes in jammed matter}

\author{Daniel M. Sussman}
\email[]{dsussman@sas.upenn.edu}
\author{Olaf Stenull}
\author{T. C. Lubensky}
\affiliation{Department of Physics and Astronomy, University of Pennsylvania, 209 South 33rd Street, Philadelphia, Pennsylvania 19104, USA}

\begin{abstract}
Granular matter at the jamming transition
is poised on the brink of mechanical stability, and hence it is
possible that these random systems have topologically protected
surface phonons. Studying two model systems for jammed matter,
we find states that exhibit distinct mechanical topological
classes, protected surface modes, and ubiquitous Weyl points.
The detailed statistics of the boundary modes shed surprising
light on the properties of the jamming critical point and help
inform a common theoretical description of the detailed
features of the transition.
\end{abstract}

\maketitle

\section{Introduction}
\footnotetext{\textit{$^{\dagger}$~DMS and OS contributed
equally to this work}}

Topological properties of operators defined as functions of
wavevectors in the Brillouin zone (BZ) control and protect
aspects of the bulk electronic spectrum and the nature of
interface states \cite{Nakahara2003,volovik03,Volovik2007} in a
wide range of systems, including polyacetylene
\cite{ssh,jackiw76}, quantum Hall materials
\cite{halperin82,haldane88}, topological insulators
\cite{km05b,bhz06,mb07,fkm07,HasanKane2010,QiZhang2011}, and
Weyl semimetals
\cite{WanVish2011,BurkovBal2011a,BurkovBal2011,LiuVan2014,XuHa2015}.
Recent work
\cite{KaneLub2014,LubenskySun2015,PauloseVit2015,PauloseVit2015-b}
has shown that topology plays a similar role in protecting
phonon spectra and interface states in ball-and-spring Maxwell
lattices \cite{footnote2}. Maxwell lattices, which include the
square, kagome, and twisted kagome \cite{SunLub2012} lattices
in two dimensions and the simple cubic and pyrochlore lattices
in three dimensions, are characterized under periodic boundary
conditions by a perfect balance between the number of
constraints and degrees of freedom, i.e., $N_B=dN$ where $N_B$
is the number of lattice bonds occupied by springs, $N$ is the
number of sites and $d$ is the spatial dimension (or
equivalently $n_B=dn$, where $n_B$ is the number of springs and
$n$ the number of sites per unit cell).

The initial work on generalized kagome lattices with $3$-site
unit cells \cite{KaneLub2014} and fully-gapped phonon spectra
(i.e., with zero modes only at wavenumber $\kv = \bm{0}$)
has been extended to $4$-site-unit-cell generalized square
lattices \cite{RocklinLub2016,Po2014} in two dimensions and
pyrocholore lattices \cite{OlafTom2015} in three dimensions,
which exhibit topologically protected zero-modes, respectively,
at points or along lines in the interior of the Brillouin zone
(BZ). We refer to the latter as Weyl points or lines because of
their strong analogy with topologically protected electronic
states in Weyl \cite{BurkovBal2011,LiuVan2014} and line-node
semi-metals \cite{BurkovBal2011a}, as well as certain photonic
crystals \cite{LuJoan2013}. In addition, both experiment and
theory have demonstrated the existence of phonon edge states
\cite{Khanikaev2015,Nash2015,Wang2015a,WangBer2015,SuesstrunkHub2015,Yang2015,Peano2015,Xiao2015,
MousaviWang2015,Kariyado2015} protected by broken time-reversal
and/or inversion symmetry much as is the case in topological
insulators. These investigations in the growing field of
topological mechanics hold out the promise of eventually
controllably tuning phononic metamaterials by marshalling their
topological features.

\begin{figure}
\centerline{\includegraphics{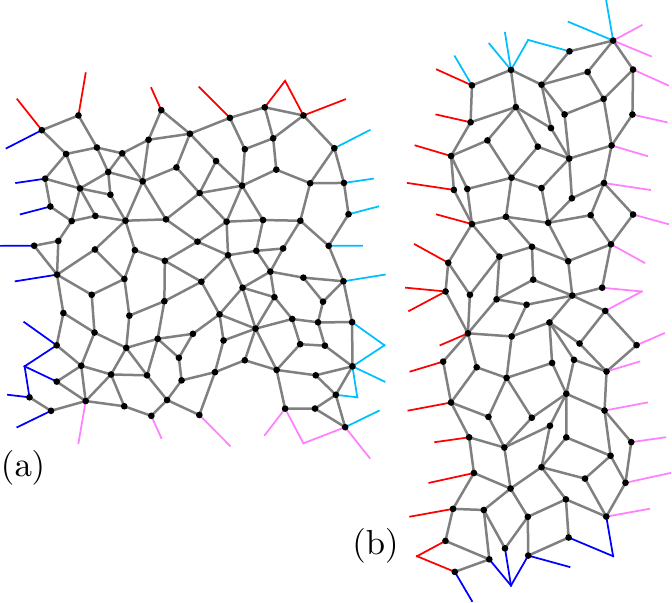}}
\caption{Representative  unit cells for (a) a jammed sphere packing (JSP)
and (b) for generic Penrose tiling (GPT).  These unit cells represent
the entire sample under periodic boundary conditions in models of jamming.
We periodically repeat them so as to create excitations with nonzero wavenumber $\kv$.
The models of jammed solids correspond the zero $\kv$ limit of the periodic
lattices. The edge bonds shown in dark red and blue are the ones that are cut to liberate the
cell from periodic boundary conditions.  The pink and light blue bonds are identical to
the dark ones under periodic boundary conditions with a single unit cell.}
\label{fig:sample_lattices}
\end{figure}

Here we apply topological ideas to large disordered-unit-cell
Maxwell lattices relevant to the jamming transition.
Topologically protected surface modes may be particularly
interesting in these systems, since the boson peak in the
density of vibrational states and diverging length scales near
the critical point of jammed systems have been explained by how
the introduction of free surfaces changes the phonon spectrum
\cite{WyartWit2005b,Wyart2005a,Sussman2015}. We study two
model ensembles related to jammed two-dimensional matter:
disordered jammed sphere packings (JSP) generated via a
compression-based algorithm~\cite{Ohern2003} and generic
periodic approximations of the Penrose tiling
(GPT)~\cite{Stenull2014,Moukarzel2015,OlafTomReply}. Figure
\ref{fig:sample_lattices} shows representative examples of
relatively small unit cells from these ensembles. The large
size of the unit cells we investigate precludes the sort of
systematic study of the full phase space of unit-cell
configurations and their associated topological properties that
has been carried out for small unit cell lattices. Instead we
pursue a stochastic approach in which we sample lattices
composed of different randomly configured, but periodically
repeated unit cells, and imagine taking the
large-unit-cell-size limit. The configurations we study almost
always have Weyl points in their spectra; that is, almost none
has a fully gapped spectrum for all nonzero wavenumbers $\kv$
in the BZ. The distribution of Weyl points in the unit cell and
their number changes with random configuration, leading to
different distributions of zero modes at free surfaces, and
thus to different topological characterizations, even at the
$\kv=\bm{0}$ point most relevant to jamming. In jammed systems,
the $n$-site unit cell is in fact the entire system (i.e.,
$n=N$). However, by studying lattices in which these unit cells
are periodically repeated \cite{Schoenholz2103}, we are able to
identify surface modes that do not appear in the bulk spectrum
and that penetrate an arbitrary number of unit cells into the
bulk. We find that the distribution of surface-mode decay
lengths undermines part of the standard cutting argument
\cite{WyartWit2005b,Wyart2005a} used to predict the density
of states and the divergent length scale $l^*$.

The remainder of the paper is organized as follows. In Section
\ref{sec:methods}, we briefly describe our protocol for
generating members of the JSP and GPT ensembles. We also detail
how the topological index and protected surfaces modes of these
lattices can be theoretically and numerically computed. Section
\ref{sec:topoModes} applies these ideas to our model disordered
systems, and demonstrates that these large-unit-cell systems
posses a correspondingly large number of topologically
protected modes including Weyl modes and edge modes both at
free surfaces and at domain walls which connect cells of
differing topological character. In Section \ref{sec:Jamming},
we present detailed statistics of the edge modes in our
systems, and discuss the implications of those statistics for the ``cutting argument\cite{WyartWit2005b,Wyart2005a}''  commonly
used to explain properties of jammed systems. We close in
Section \ref{sec:disc} with a brief summary.

\section{Models and methods\label{sec:methods}}
\subsection{Construction of lattices}
We produce our model networks as follows: For the JSP, we
numerically generate packings of $N$ particles in two
dimensions by starting with a dense (i.e., above the jamming
transition) polydisperse mixture of discs with a flat
distribution of particle sizes between $\sigma$ and
$1.4\sigma$, where $\sigma$ is the unit of length. The
interaction between particles is modeled by a harmonic soft
repulsive potential~\cite{Ohern2003}. The discs are initially
placed at random (i.e. in an infinite-temperature
configuration) in a square simulation box with periodic
boundary conditions. The system is then rapidly quenched to
zero temperature by combining line-search methods, Newton's
method, and the FIRE algorithm \cite{quench}. The average
coordination number is changed by incrementally expanding or
compressing the system uniformly and then re-quenching to zero
temperature; this is repeated until $n_B=dn$, that is, until we
obtain a Maxwell lattice. The soft disc packing is converted to
the ``unstressed" network \cite{Alexander1998, Silbert2005pre},
replacing each pairwise interaction with an un-stretched
harmonic spring between nodes at the particle centers.

For the GPT, we use the standard projection procedure from the
five-dimensional hypercube $\mathbb{Z}^5$ onto a
two-dimensional space \cite{Bruijn1981}. The orientation of the
plane in hyperspace that leads to the quasiperiodic rhombus
tiling~\cite{Penrose1974} is related to the the golden ratio
$\tau$. Approximating $\tau$ by the ratios $\tau_m$ of
successive Fibonacci numbers ($\tau_1 = 1/1$, $\tau_2=2/1$,
$\tau_3=3/2,\ldots$) gives the periodic approximants. These are
rhombic tiles arranged in rectangular unit cells of increasing
size that approach the quasiperiodic tiling as
$m\rightarrow\infty$. We randomly displace nodes by a small
amount without changing the connectivity of the approximants,
and then replace the edges of the (deformed) tiles with
un-stretched harmonic springs.

\subsection{Edge states and topological characterization\label{sec:edge-top}}
The vibrational properties of elastic networks consisting of
periodically repeated unit cells with $n$ sites and $n_B$ bonds
can be described \cite{Calladine1978} by the $n_B \times dn$
compatibility matrix $\brm{C}(\kv)$, relating bond
displacements $\uv(\kv)$ to bond extensions $\ev(\kv)$ via
$\brm{C}(\kv) \uv(\kv) = \ev(\kv)$, and the $dn\times n_B$
equilibrium matrix $\brm{Q}=\Cv^\dag (\kv)$, relating bond
tensions $\tv(\kv)$ to site forces $\fv(\kv)$, for each
wavenumber $\kv$ in the BZ. The null space of $\Cv(\kv)$
consists of zero modes whose displacements do not stretch
bonds; that of $\Qv(\kv)$ consists of states of self stress
(SSS) in which bonds under tension exert no net forces at
sites. When masses and spring constants are all set to unity,
the dynamical matrix determining the phonon spectrum is simply
$\Dv(\kv) = \Qv(\kv) \Cv(\kv)$. In periodic systems, the
Calladine-Maxwell theorem \cite{Calladine1978,LubenskySun2015}
generalizes to $n_0(\kv) - s(\kv) = dn - n_B$ for every $\kv$,
where $n_0(\kv)$ is the number of zero modes and $s(\kv)$ the
number of SSS. In periodic Maxwell lattices where $dn=n_B$
there is always one SSS for each zero mode.

The term ``isostatic'' rather than ``Maxwell'' is commonly used
to identify lattices with $N_B=dN$.  This term, however,
strictly speaking only applies to finite lattices (frames)
\cite{Carpinteri1997,ConnellyWhi2009} with  no SSSs and only
zero modes corresponding to the $d(d+1)/2$ rigid translations
and rotations $d$-dimensions, i.e. to ``statically
determinate'' lattices. Under periodic boundary conditions,
things are complicated in $N_B=dN$ lattices by the presence of
zero energy elastic distortions of the unit cell, called Guest
modes \cite{GuestHut2003,GoodrichNag2014,LubenskySun2015}, but
it is reasonable to use the term ``isostatic'' for Maxwell
lattices ($n_B = dn$) with exactly $d$ zero modes and $d$ SSSs.
Thus periodic isostatic lattices are a subset of periodic
Maxwell lattices, and Maxwell lattices with Weyl modes are not
isostatic.

\subsubsection{Winding numbers and domain walls}\hspace*{\fill}

\begin{figure}
\centerline{\includegraphics{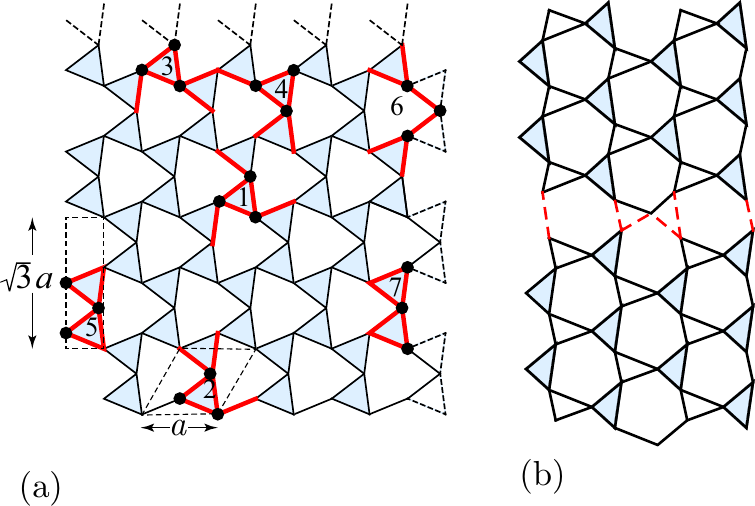}}
\caption{(a) A section of a twisted kagome lattice showing
		dashed bonds at the top and right edges that are cut to liberate the multi-unit cell lattice from periodic boundary conditions on a torus. Also shown are different unit cells, including a symmetric
		unit cell (1) that is not compatible with the surfaces and various
		surface compatible unit cells (2-7). (b) An example of a domain
		wall, whose bonds are dashed, connecting two lattices with
		different topological polarization.}
	\label{fig:unit_cells}
\end{figure}

The determinant of $\Cv(\kv)$ (or $\Qv(\kv)$) provides a map
from the BZ to the complex plane. Any path in wavenumber space
starting and ending at points separated by a reciprocal lattice
vector $\Gv$ is mapped to a closed loop in the complex plane
characterized by a winding number. For simplicity, we consider
straight paths parallel to reciprocal lattice vectors indexing
sets of lattice planes perpendicular to them. Let $p$ be the
component of $\kv$ parallel to $\Gv$ and $q$ the component
parallel to these planes. Then $\det \Cv(\qv,\Gv) \equiv \det
\Cv(q,p,\Gv)$ depends on $p$ through $z= e^{i 2 \pi p/G}$,
where $G = |\Gv|$. The winding numbers,
\begin{equation}
n(q,\Gv) = \frac{1}{2 \pi i} \int_0^G dp \frac{d}{dp} \ln \det
\Cv(q,p,\Gv) ,
\label{eq:windNum}
\end{equation}
depend in general on $q$ along a surface as well as $\Gv$. In the small-unit-cell systems previously considered this integral either vanished or gave $\pm 1$, but in general it can take any integer value bounded by the number of bonds cut by the surface.
The matrices $\Cv(\kv)$ and $\det \Cv(\kv)$ also depend on the choice of
unit cell as depicted in Fig.~\ref{fig:unit_cells} (a). If the
cell is chosen so that it is ``surface compatible", i.e., so
that it has no nodes or ``dangling" bonds outside a lattice bounded by a
surface coinciding with a lattice plane indexed by $\Gv$, then the local contribution to the surface zero mode count vanishes and
the winding number computed by Eq. (\ref{eq:windNum}) gives the
number of zero modes localized on that surface: $n_0^S(q, \Gv)
= n(q,\Gv) \geq 0$. If a portion of the lattice under periodic boundary conditions is liberated by cutting two parallel free surfaces that remove $\Delta
n_B$ bonds (for simplicity, we do not consider removal of sites as
well), then the total number of zero modes on the two surfaces is
$n_{0,\rm{tot}}=n_0^S ( q ,\Gv) + n_0^S(q, -\Gv)=\Delta n_B$.

At a domain wall separating ``left'' and ``right'' lattices,
the $q\neq0$ topological count \cite{LubenskySun2015} of the
difference between the number of domain-wall zero modes,
$n_0^D(q,\Gv)$, and SSS, $s^D(q,\Gv)$, is $\nu^D(q,\Gv) \equiv
n_0^D (q,\Gv) -s^D(q,\Gv) = n_{0-}^S (q,\Gv) + n_{0+}^S
(q,-\Gv) - \Delta n_B$, where $n_{0,\mp}^S (q,\pm \Gv)$ are the
number of zero modes of the left and right free surfaces that
will constitute the domain wall and $\Delta n_B$ is the number
of bonds per unit cell needed to bind the two free surfaces
together. Domain walls prepared in this way can have either
zero modes ($\nu^D(q,\Gv) =n_0^D (q, \Gv)$) or SSS
($\nu^D(q,\Gv) = -s^D(q,\Gv)$). The topological properties of
fully-gapped Maxwell $2d$-lattices, like the kagome and some
realizations of the $4$-site-unit-cell square lattice, are
fully determined by a polarization vector $\Rv_T = -\sum n(q,
\bv_i) \av_i$, where $\av_i$ and $\bv_i$ are the basis vectors
of the direct and reciprocal lattice, respectively, and
$n(q,\bv_i)$ is independent of $q$. A useful but $q$-dependent
measure of topological character in systems with Weyl points is
$\nu^W (q,\Gv) = [n_0^S(q,\Gv) - n_0^S(q,-\Gv)]/2$, which
reduces to $\Rv_T\cdot \Gv/(2 \pi)$ in gapless systems in
agreement with reference \cite{KaneLub2014}.

Since the $q=0$ limit is of particular interest in jammed
systems, we note that when $q=0$ the topological count at a
domain wall is slightly modified by the presence of the global
translational zero modes. The counting of the number of zero
modes at a free surface includes these global translational
modes, relative rigid translations of the cells on either side of the domain wall, and the $\tilde{n}_{0,\pm}^S(0,\Gv)$ of
exponentially decaying surface modes. Thus, there are a total of
$\tilde{n}_{0,\pm}^S(0,\Gv) + d$ zero modes at the $\pm$
surfaces. As when $q\neq 0$, the creation of the domain wall
requires $\Delta n_B$ extra constraints, and the total number
of domain-wall zero modes (excluding global translations is)
\begin{align}
\tilde{\nu}^D(0,\Gv) & =\tilde{n}_{0,+}^S(0,\Gv) + d+
\tilde{n}_{0,-}^S(0,\Gv) + d-\Delta n_B - d \nonumber\\
& = \tilde{n}_{0,+}^S(0,\Gv) +\tilde{n}_{0,-}^S(0,\Gv) + d-\Delta n_B .
\end{align}
This count includes $d$ modes in which the $+$ and $-$ surfaces
translate rigidly with respect to each other.  These modes can
in general mix with the exponentially decaying modes.

\subsubsection{Structure of surface modes}\hspace*{\fill}

The winding numbers computed via Eq.~(\ref{eq:windNum}) provide
a count of the number of zero modes at a free surface. Much
more information about these modes can be extracted from the
compatibility matrix.  The lattice can be divided into
contiguous layers, one unit cell thick and composed of parallel
contiguous surface-compatible unit cells starting at the free
surface and penetrating inward. This construction allows the
compatibility matrix to be Fourier transformed parallel to the
layers to produce a banded matrix $\brm{C}(q,\Gv)$ \cite{LubenskySun2015}. The main diagonal of this banded matrix is composed of $dn\times n_B$ submatrices $\brm{C}_{11}(q,\Gv)$; the $\brm{C}_{11}$ are compatibility matrices that describe the relationship of node motions to bond extensions for nodes and bonds entirely within a layer. These are indicated, e.g., by the gray and blue bonds in Fig.
\ref{fig:sample_lattices}. The diagonal in $\brm{C}$ above the main diagonal is composed of identically-sized submatrices $\brm{C}_{12}(q,\Gv)$, which connect bonds in one layer to sites in the next layer, indicated by the red bonds in Fig. \ref{fig:sample_lattices}. For a fully periodic system with no surface, by construction $\brm{C}(q) = \brm{C}_{11}(q)+\brm{C}_{12}(q)$.

In this
construction, a set of displacements $\brm{U} =\left(
\vec{u}_1,\vec{u}_2,\ldots \right)$, where $\vec{u}_i$ is a set
of displacements in unit cell $i$, will be a zero mode if
$\brm{C}_{11} \vec{u}_i+\brm{C}_{12}\vec{u}_{i+1} = 0.$ These
equations, in turn, are solved by $\vec{u}_{i+1} = \lambda
\vec{u}_i$ if
\begin{equation}
\label{eq:det1}
\det \left(\brm{C}_{11} + \lambda \brm{C}_{12} \right) =0 ,
\end{equation}
and modes decay as $\lambda = \exp (-\kappa r)$ with distance
$r$ (in units of the unit-cell size) away from the free surface
when $\lambda <1$. In general, the inverse penetration depth
$\kappa$ is complex, indicating a surface mode that decays
exponentially with oscillations. The sign of $\tilde{\kappa} =
\rm{Re}(\kappa )$ determines which surface the zero mode is
localized to: positive (negative) $\tilde{\kappa}$ goes with
the surface bounding an interior toward positive (negative)
$r$.

For small-$n$ unit cells this prescription works well. In
general, though, evaluating the determinant of a large, sparse
matrix and finding the roots of the resulting polynomial in
$\lambda$ is both slow and numerically unstable, making it
difficult to find all of the localized surface modes. However,
if $\brm{C}_{11}$ is an invertible matrix (implying no zero
mode localized completely in the surface unit cell), the
problem can be reduced to finding the eigenvalues $S_j$ of
$\brm{S}\equiv \brm{C}_{11}^{-1}\brm{C}_{12}$: the $\lambda$'s
that satisfy Eq. (\ref{eq:det1}) are determined by the set of
non-zero $S_j$, $\lambda_j = -S_j^{-1}$. For the GPT up to the
$8/5$ approximant ($n=1440$), we were always able to choose a
unit cell with an invertible $\brm{C}_{11}$. For jammed unit
cells it becomes increasingly hard with increasing $n$ to find
unit cells with an invertible $\brm{C}_{11}$, implying the
existence of zero modes completely localized within the unit
cell adjacent to the free surface. This is not surprising in
light of existing data on the prevalence of surface
``rattlers'' in the presence of cut surfaces \cite{Wyart2005}.
Nevertheless, in Section \ref{sec:Jamming} we will show that
these surface rattlers do not seem to affect the distribution
of the other surface modes of the system.

\section{Topological modes of disordered systems\label{sec:topoModes}}
\subsection{Surface and Weyl modes}
\begin{figure}[!ht]
\centerline{\includegraphics[width=1.\linewidth]{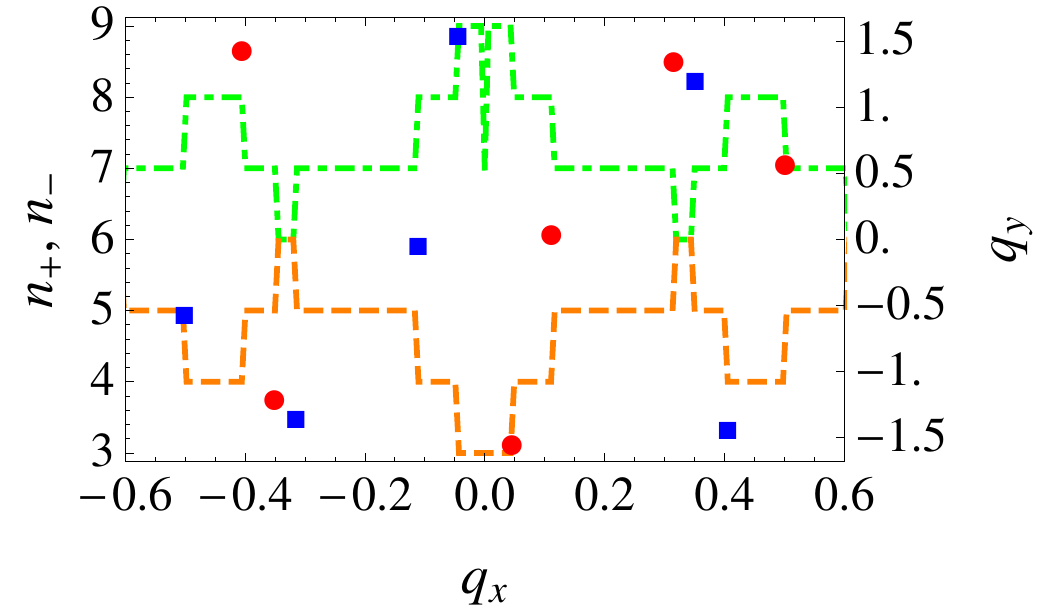}}
\caption{\label{fig:windvsurf} (Color online) Number $n_+$ of
surface zero modes with positive $\tilde{\kappa}$ (green dot-dashed line)
and number $n_-$ with negative $\tilde{\kappa}$ (orange dashed
line) for a typical $1/1$ approximant (which has 30 sites per unit cell) as a function of $q_x$. Square (blue) and
circular (red) points indicate the location (with $q_y$ on the
right ordinate) of Weyl points with positive and negative
charge, respectively. The uniform translation modes at $q=0$
are not included, and hence $n_+ + n_-$ decreases by $d=2$ at $q=0$.
The topological count is $\tilde{\nu}^W(0,\Gv )
=(7-3)/2 = 2$.}
\end{figure}

Figure~\ref{fig:windvsurf} illustrates a topological
characterization for a typical realization of the $1/1$
approximant of the GPT. The plot shows the number $n_+$ ($n_-$)
of surface zero modes with positive (negative) $\tilde{\kappa}$
as a function of wavevector. The discontinuous jumps in the
topological class as a function of $q=q_x$ imply the presence
of Weyl-type singularities. To locate these singularities, we
calculate the winding numbers (given by Eq.~(\ref{eq:windNum}),
but for small closed square loops instead of straight
integration lines) around points in the BZ. These integrals
yield $0$, $-1$ or $+1$, indicating the absence or,
respectively, presence of a Weyl point with charge $-1$ or
$+1$. Here, there are $N_{\text{Weyl}}=12$ Weyl points that
come in pairs of opposite charge at positions $\brm{k}$ and $-
\brm{k}$. Discontinuous changes in $n_+$ and $n_-=\Delta
n_B-n_+$ occur at projections of these points onto the $q_x$
axis and have a magnitude equal to the winding number of the of
the Weyl point. We note that for this example $n_+(q=0) \neq
n_-(q=0)$. This indicates that, in the $q=0$ limit of
particular relevance for jammed systems, we can find states
with non-trivial topological character.

\begin{figure}[!ht]
\centerline{\includegraphics[width=1.\linewidth]{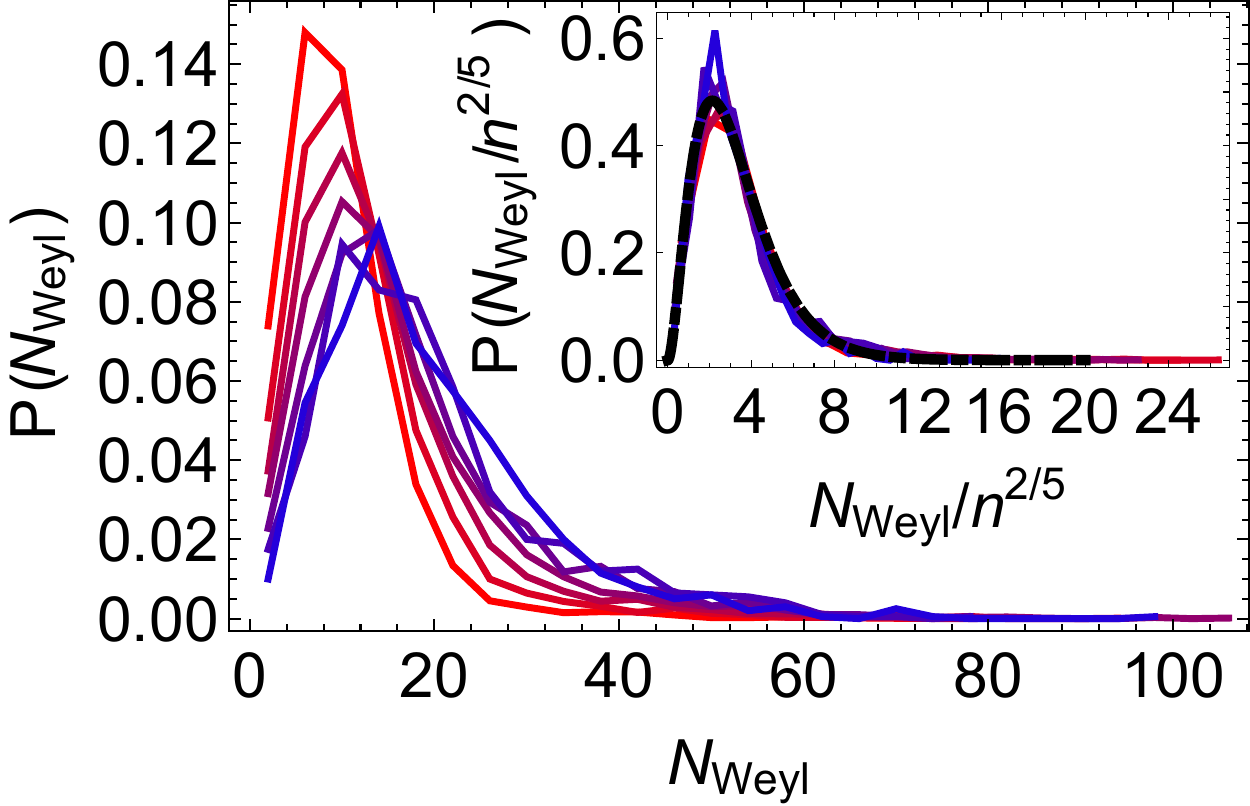}}
\caption{\label{fig:weyldist} (Color online) Probability distribution of the number of
Weyl points for JSP with $n=16,\ 24,\ 32,\ 48,\ 64,\ 80,\ 96$
(red, narrowly peaked to blue, broad curves). The inset shows the curves
in the main frame scaled by $n^{2/5}$, together with a fit to a gamma
distribution with parameters $k=2.83$ and $\theta=0.582$ (black, dashed curve).}
\end{figure}

In general we find that our disordered systems have an
increasing number of Weyl points as the size of the unit cell
grows. This data is shown in Fig.~\ref{fig:weyldist} for the
JSP ensemble and unit cell sizes between $n=16$ and $n=96$.
Both the mean and the distribution of the number of Weyl points
in a given realization of the disordered system scales
approximately as $n^{2/5}$, although given the discrete nature
of the distribution exponents in the range $1/3$ to $1/2$
collapse the data in the figure almost as well. For states in
the GPT ensemble the scaling is similar. Reference \citenum{Schoenholz2103} carefully studied the negative eigenvalues that sometimes appear at finite wavevectors in hyperstatic sphere packings under pressure, and attributed these instabilities to stress. In our systems, applying a small pressure term while fixing the contact network would generically push the Weyl modes we have found to negative frequencies. Thus, the question of whether the Weyl points in this work and the stress-induced instabilities of Ref.\citenum{Schoenholz2103} are independent and complementary instabilities or if they are linked to each other away from the Maxwell lattice limit is an interesting open question. Also interesting is whether any of the Weyl modes we have found here survive in mildly hyperstatic but unstressed packings.

\subsection{Domain-wall modes}
\begin{figure}[!ht]
\centerline{\includegraphics[width=1.0\linewidth]{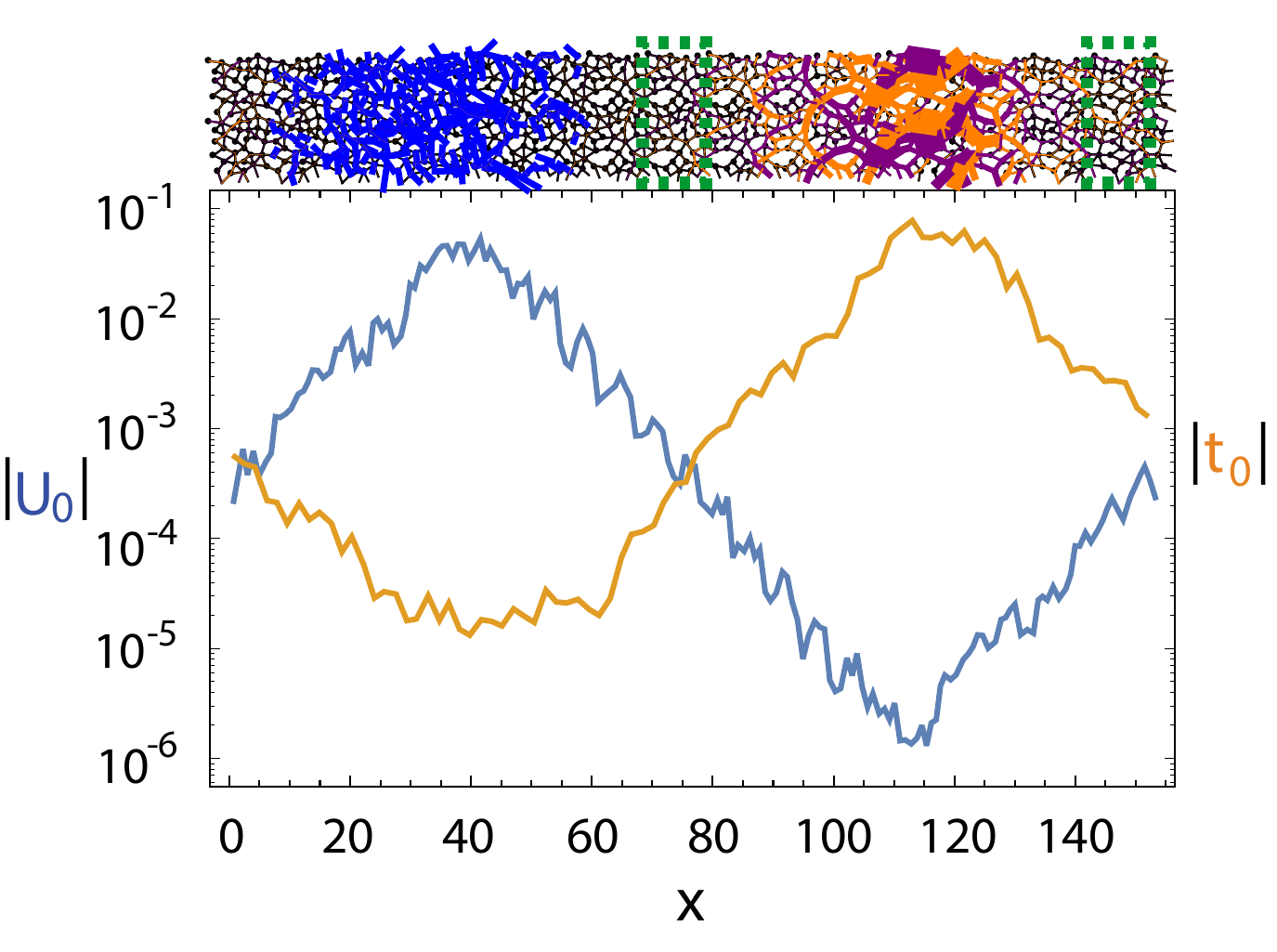}}
\caption{\label{fig:topoglue} (top) State of self stress and
zero mode localized at the interface between jammed unit cells
with $n=64$ of different topological classes. The large,
rectangular unit cell is under periodic boundary conditions in
both directions. Dashed boxes highlight the unit cells that are repeated and glued together to from the supercell. (bottom) Total magnitude of the topologically
protected zero mode (blue) and state of self stress (yellow) in
vertical slices of the combined unit cell. The exponentially
localized character of each, with oscillations, is clearly
seen.}
\end{figure}

To study topologically protected phonons localized at domain
walls separating different topological classes, we construct
Maxwell ``supercells" by joining two lattices with unit cells
$A$ and $B$ -- which have the same number of boundary bonds --
in the sequence $A\cdots AB\cdots BA\cdots A$. Unit cells at
the $AB$ and $BA$ interfaces are linked with the appropriate
number of bonds to preserve the Maxwell relation $dn=n_B$ when
periodic boundary conditions are applied to the outer $A$'s,
but otherwise the linking bonds are arbitrary. In
Fig.~\ref{fig:topoglue} we show a representative example for a
JSP lattice with a total of $10 A$ and $10 B$ cells, each with
$n=64$ sites, and focus on calculating the normal modes of the
system at $q=0$. We chose the $A$ and $B$ lattices to be in
different topological classes, according to the value of the
winding number calculated with Eq.~(\ref{eq:windNum}). In an
infinite system, one would, therefore, expect that one of the
domain walls would exhibit zero modes and the other SSS. In our
finite system, interaction between these two domain walls raise
the frequency of zero modes to (exponentially small) nonzero
values. And indeed, a direct numerical evaluation of the
eigenmodes of the dynamical matrix (and the dual matrix $
\mathcal{M}(\kv)=\Cv(\kv=0)\Qv(\kv=0)$) reveal that there is a
zero mode (in the infinite size limit) at one of the two
interfaces, and as demanded by the Calladine-Maxwell
count~\cite{Calladine1978}, a balancing SSS which is located at
the other.

\begin{figure}[!bhp]
\centerline{\includegraphics[width=0.95\linewidth]{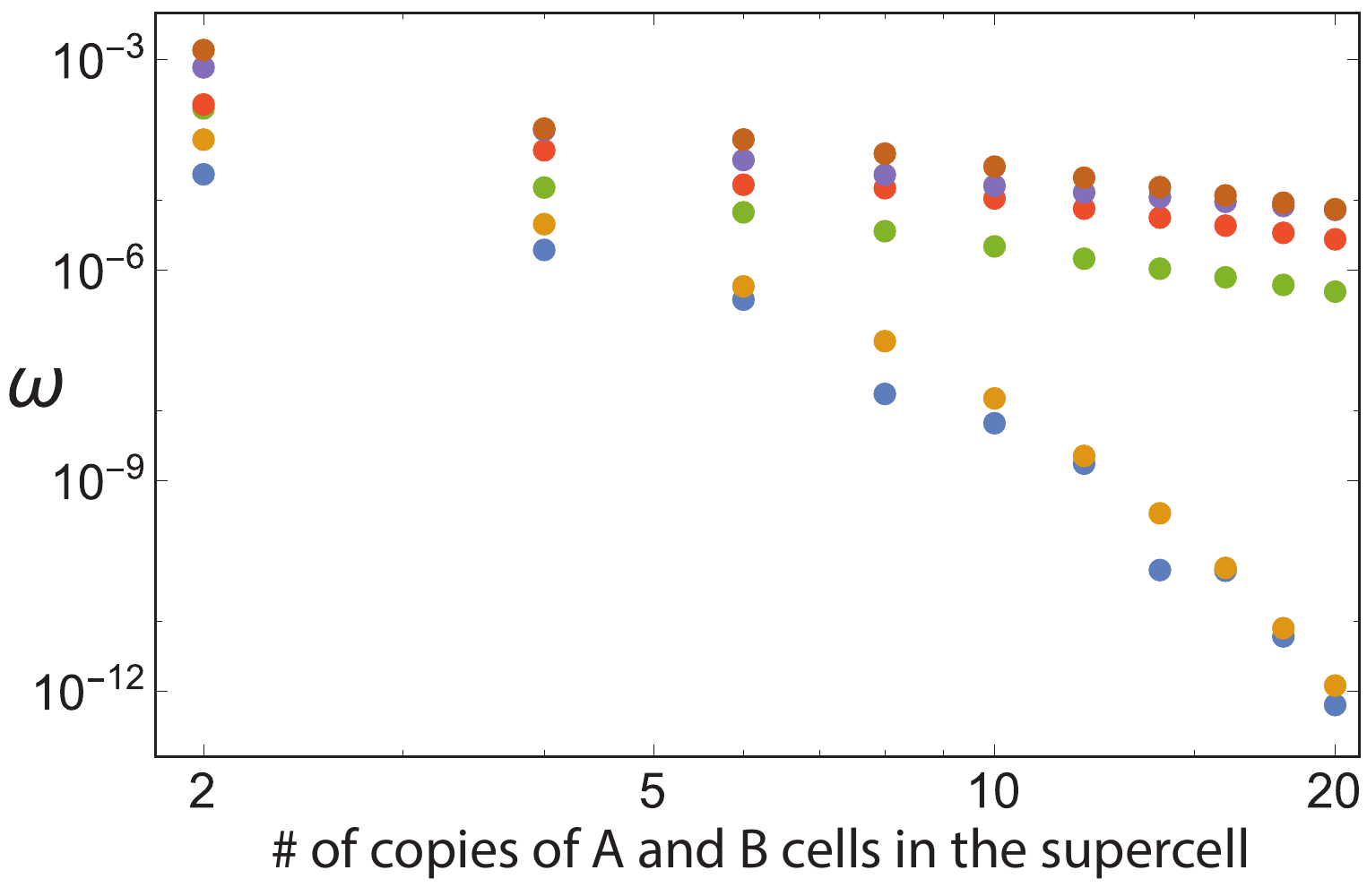}}
\caption{\label{fig:wdrop} Frequencies (square-root of the energetic cost) of the lowest
vibrational modes at the interface of jammed unit cells of
different topological classes as a function of how many times
the unit cells were copied. Here we see two modes with
exponentially decaying frequency, together with four disordered plane waves modes with
a power-law decay.}
\end{figure}

\begin{figure}[!htbp]
\centerline{\includegraphics[width=1.\linewidth]{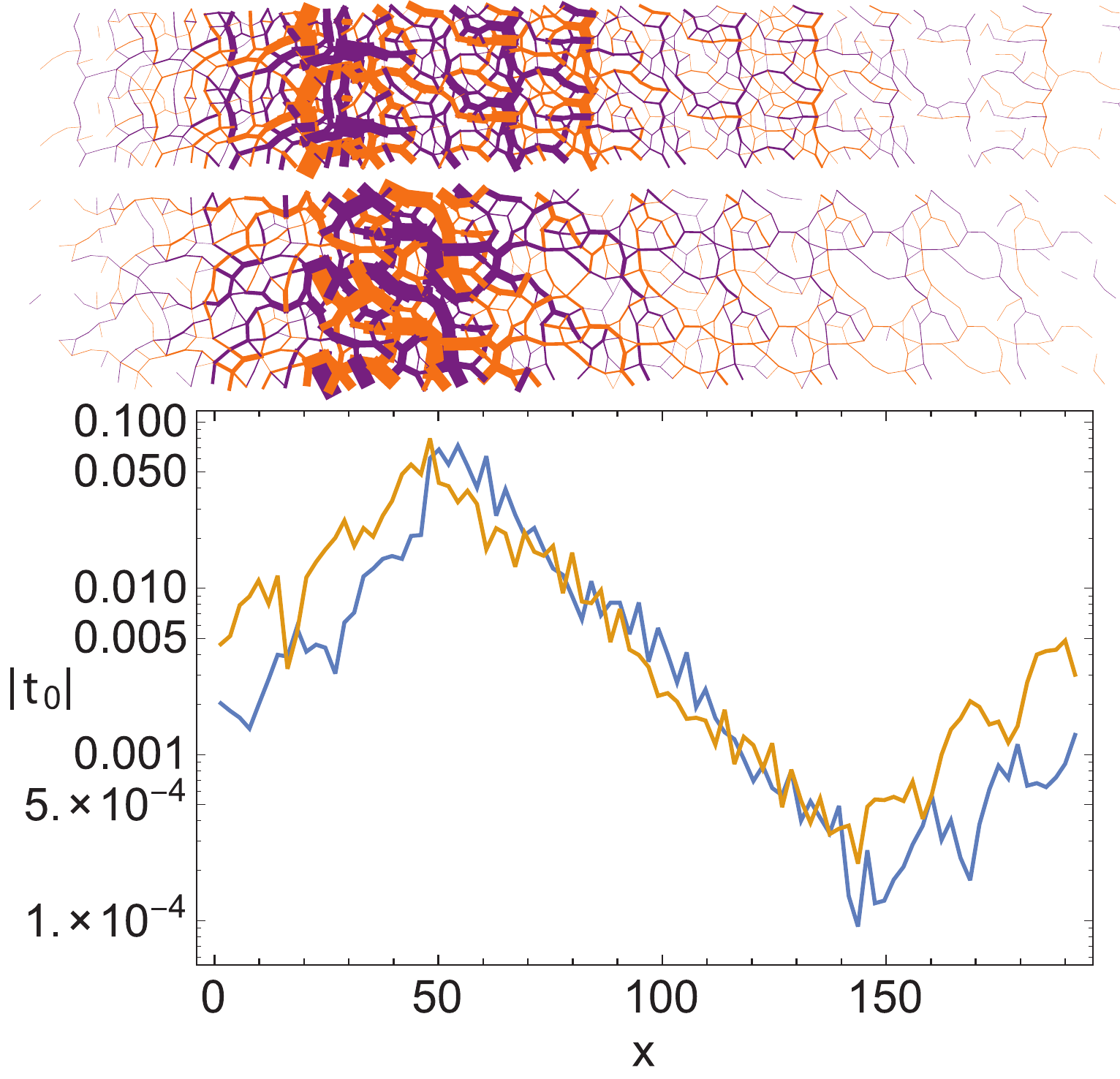}}
\caption{\label{fig:topoglue2} (top) Two independent states of
self stress localized at the interface between tiled jammed
unit cells of $n=24$ with $\nu^d(0,\bm{y})=2$. For clarity,
bonds whose stress is below a threshold are not shown.}
\end{figure}

As discussed above, there can be more that one zero mode per
wavenumber $q$ along a domain wall, since it is the combination
$ n_{0-}^S (0,\Gv) + n_{0+}^S (0,-\Gv) - \Delta n_B$ that
determines the number of interfacial zero modes or SSS at a
domain wall. By carefully selecting different unit cells on
either side of a domain wall it should be possible to create an
interface that hosts multiple topologically protected modes
localized to the domain wall. We present such an example here.
As noted above, the exponentially decaying surfaces modes will
have a small, non-zero frequency when they are inserted into
the finite $A\cdots AB\cdots BA\cdots A$ supercells we use
(related to the magnitude of the exponentially decaying mode
when it encounters the next interface). One way of detecting
them, then, is to look for modes whose frequencies become
exponentially small as more and more copies of $A$ and $B$ are
used in the construction of the supercell. This is shown in
Fig.~ \ref{fig:wdrop}. While many of the modes decay as a power
law with increasing system size and are thus easily identified
as being low-frequency disordered plane waves, we see two
modes whose energy scale drops exponentially fast. This is a
clear signature of modes exponentially localized to an
interface in the supercell. As illustrated in
Fig.~\ref{fig:topoglue2}, which shows the bonds participating
in the two distinct SSSs, these multiple interfacial modes can
readily be visualized. There are of course compensating domain
walls with zero modes, not shown in the figure.

\section{Statistics of Surface Modes\label{sec:Jamming}}
\subsection{Distribution of penetration depths}
\begin{figure}
\centerline{\includegraphics[width=1.0\linewidth]{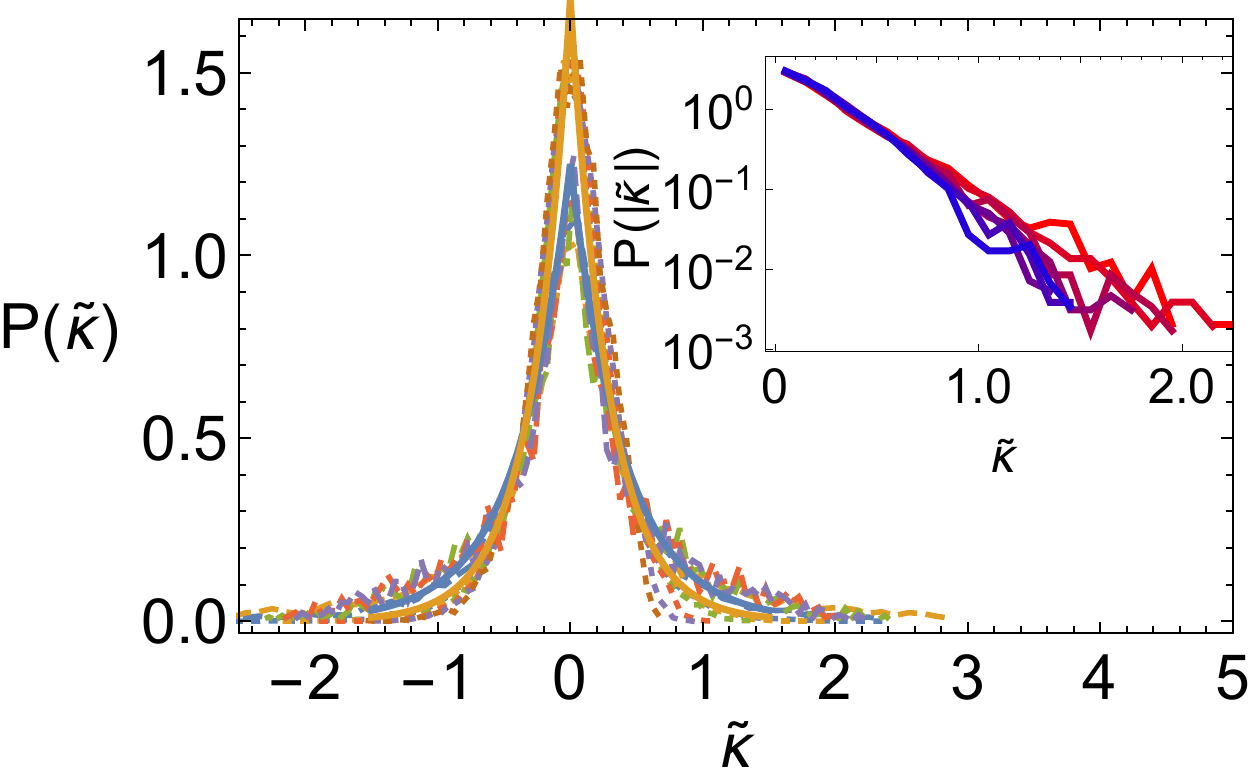}}
\caption{\label{fig:khist} (Color online)  Probability
distribution of $\tilde{\kappa}$ for (dotted lines) JSP with
$n=16,\ 24,\ 32,\ 48,\ 64,\ 80,\ 96,\ 128$ and (dashed lines)
GPT approximants $1/1,\ 2/1,\ 3/2,\ 5/3,\ 8/5$, corresponding to $n=30, 80, 210, 550, 1440$.
Solid lines are best exponential fits to
$P(\tilde{\kappa}) \approx\exp(-|\alpha \tilde{\kappa}|)\alpha/2$ for $\alpha = 2.5$ and
$\alpha = 3.4$ for the GPT and JSP, respectively,for all  data points
independent of lattice size. (inset)  Log plot of $P(|\tilde{\kappa}|)$
for JSP with $n=16$ (top, red curve) to $n=96$ (bottom, blue curve). Note the small but noticeable increase in decay rate with increasing system size.}
\end{figure}
Now, we turn to the statistics of surface zero modes at $q=0$
for surfaces parallel to the $x$-axis. In GPTs this corresponds
to the $x$ direction of the undistorted Penrose tiling and in
JSP to one of the faces of the simulation cell (the JSP have no
unit-cell anisotropy, so the distinction between $x$ and $y$ is
unimportant). We place no restriction on depth perpendicular to
the $x$-axis, so we can access penetration lengths that span an
arbitrary number of unit cells into the bulk, and we average
over many random realizations of our model systems (e.g.,
$1000$ for the 1/1 GPT). The distributions
$P(n_+)$ and $P(n_-)$ of $n_+$ and $n_-$ for both JSP and GPT
are approximately Gaussian with mean given by half the number
of bonds that are cut to produce the free surface ($\sim
\sqrt{n}$) and a variance that grows as $\sim n^{1/4}$.
Figure~\ref{fig:khist} shows the probability distributions of
the inverse penetration depths, $P(\tilde{\kappa})$, omitting
the delta-function spike at $\tilde{\kappa} = 0$ associated
with the trivial translations. Both associated
$P(\tilde{\kappa})$s can be approximated by exponential
distributions, albeit with different variances. The
distributions for the JSP and GPT are characterized by a
slightly different expected penetration depth, and the main
plot suggests that the typical decay lengths do not grow
strongly with system size. However, zooming into the tails of
the distribution suggests that a very modest system-size effect
may be present, a point we discuss later.

\subsection{Effect of surface rattlers on JSP statistics}
\begin{figure}[!ht]
\centerline{\includegraphics[width=0.85\linewidth]{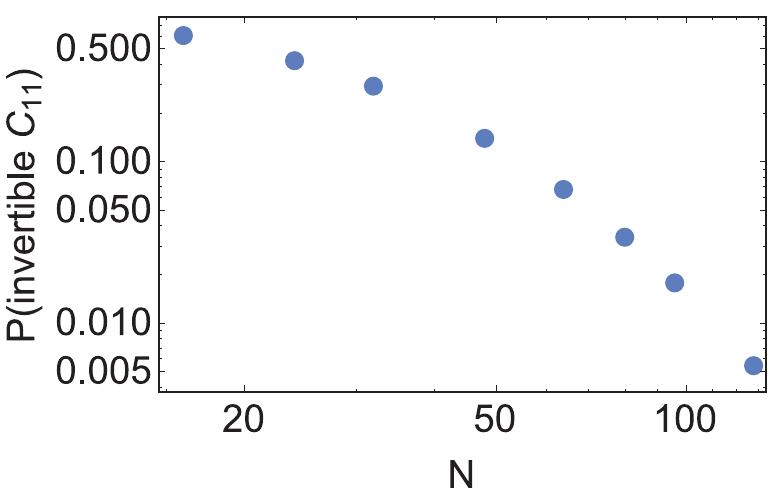}}
\caption{\label{fig:goodconfs} Probability that a random compression-algorithm-generated jammed configuration with $N$ sites will have a
choice of square unit cell for which $\brm{C}_{11}$ is
invertible.}
\end{figure}

In the previous sections, we studied lattices with square unit
cells with invertible $\brm{C}_{11}$ matrices that necessarily
had no surface zero modes with nonzero amplitude restricted to
the first layer of the lattice because if they did, there would
have to be a displacement vector $\uv$ such that $\brm{C}_{11}
\uv = 0$ contradicting the assumption that $\brm{C}_{11}$ is
invertible.  However, systems at the jamming threshold
typically have surface rattlers when free surfaces are
introduced, and the probability of finding a configuration with
an invertible $\brm{C}_{11}$ decays rapidly as the system size
increases. In Fig. \ref{fig:goodconfs}, we take a
representative subset of JSP with system size $N$ generated by
the compression algorithm, which we denote $JSP_N$, and plot
the fraction of these states with an invertible $\brm{C}_{11}$.
For those states with a non-invertible $\brm{C}_{11}$, some
modes exist that are necessarily localized entirely within the
surface unit cell but they may, nevertheless, involve particles
far from the free surface. Note, though, that the machinery
developed in Section \ref{sec:methods} can still be used
(albeit by directly solving $\det \left(\brm{C}_{11} + \lambda
\brm{C}_{12} \right) =0$ rather than using the eigenvalue
techniques) to investigate penetration depth statistics, and
noting that when $\brm{C}_{11}$ is not invertible one must be
careful to include not only the topological count discussed in
this work but also a local zero-mode count of the surface cell
relative to that of the bulk cell (this is captured by the
``dipole moment'' $\Rv_L$, which allows one to express the
local zero-mode count via $\Rv_L\cdot \Gv/(2 \pi)$, as
discussed in Refs.\cite{KaneLub2014,LubenskySun2015})

\begin{figure}[!htbp]
\centerline{\includegraphics[width=0.95\linewidth]{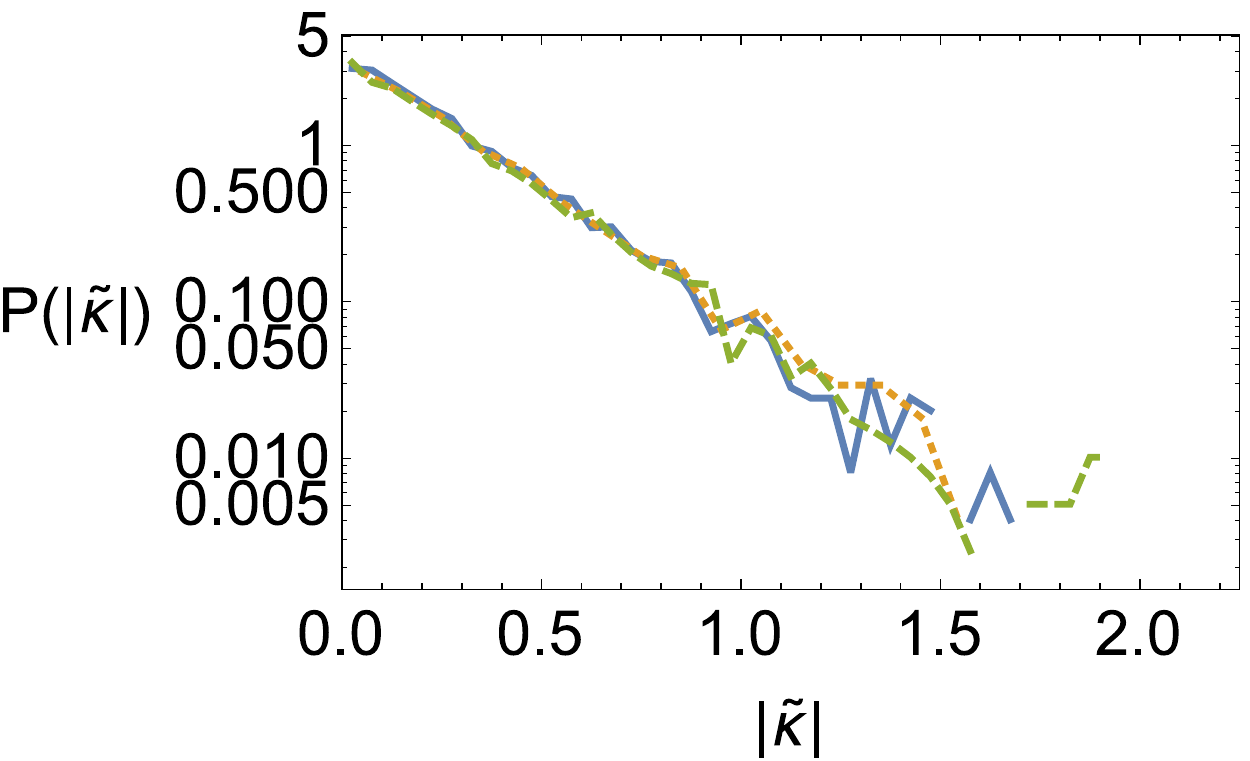}}
\caption{\label{fig:detveig} Probability distribution of
penetration depths for different sampling and numerical methods
for $n=32$ JSP. The blue solid and yellow dotted lines use the eigenvalue and determinant methods, respectively, of determining $\tilde{\kappa}$ for JSP for
which $\brm{C}_{11}$ is invertible. The dashed green line uses
the determinant method and averages over JSP regardless of
whether $\brm{C}_{11}$ is singular.}
\end{figure}

To check whether our results are sensitive to sampling over
only those jammed unit cells that have an invertible
$\brm{C}_{11}$, we compare the distribution of inverse
penetration depths, $P(\tilde{\kappa})$, for both $JSP_N$ and
$\{JSP_N | \exists \brm{C}_{11}^{-1}\}$. To do this we chose an
$n$ small enough that $\det \left( \brm{C}_{11} + \lambda
\brm{C}_{12} \right) =0$ can be reliably solved numerically to
collect penetration depth statistics for any member of
$\{JSP_N\}$. Figure \ref{fig:detveig} shows the result of this
for $n=32$, leading us to expect that the results presented
elsewhere in this paper are not biased by averaging over the
sub-ensemble of JSP for which $C_{11}$ is invertible.

\subsection{The overlap function}
Our explicit decomposition of the nullspace of $\brm{C}$ into
its constitutive surface modes provides insight into the
physics of the jamming transition. One common understanding of
the origin of the plateau in the density of states near the
transition proceeds via a variational argument on the behavior
of the eigenvectors of the dynamical matrix $\brm{D}$ when
periodic boundary conditions are replaced by free boundaries
\cite{Wyart2005}. This standard argument relies on an
assumption about the total structure of the nullspace of the
free-surface system. Assuming that boundaries normal to $x$ at
$x=0$ and $x=L_x$ are replaced by free surfaces, this
assumption is conveniently written in terms of the overlap
function \cite{Wyart2005a}:
\begin{equation}\label{eq:overlap}
f(x)dx = n_{0,\text{tot}}^{-1}
\sum_{\beta=1}^{n_{0,\text{tot}}}\sum_{x_i\in [x,x+dx]} |
\brm{u}_{i,\beta} |^2.
\end{equation}
where $0\leq x <L_x$, $\beta$ labels the
$n_{0,\text{tot}}=\Delta n_B$ zero modes, $\brm{u}_{i,\beta}$
denotes the displacement of site $i$ in mode $\beta$, and $x_i$
is the $x$ coordinate of the reference position of site $i$.
The original variational argument for configurations at the
jamming threshold assumed that $H_1 = \textrm{min}\left[ L_x
f(x) \right]$ is bounded from below by a constant that is
independent of linear system size \cite{Wyart2005,
Wyart2005a}.

\begin{figure}[!ht]
\centerline{\includegraphics[width=0.9\linewidth]{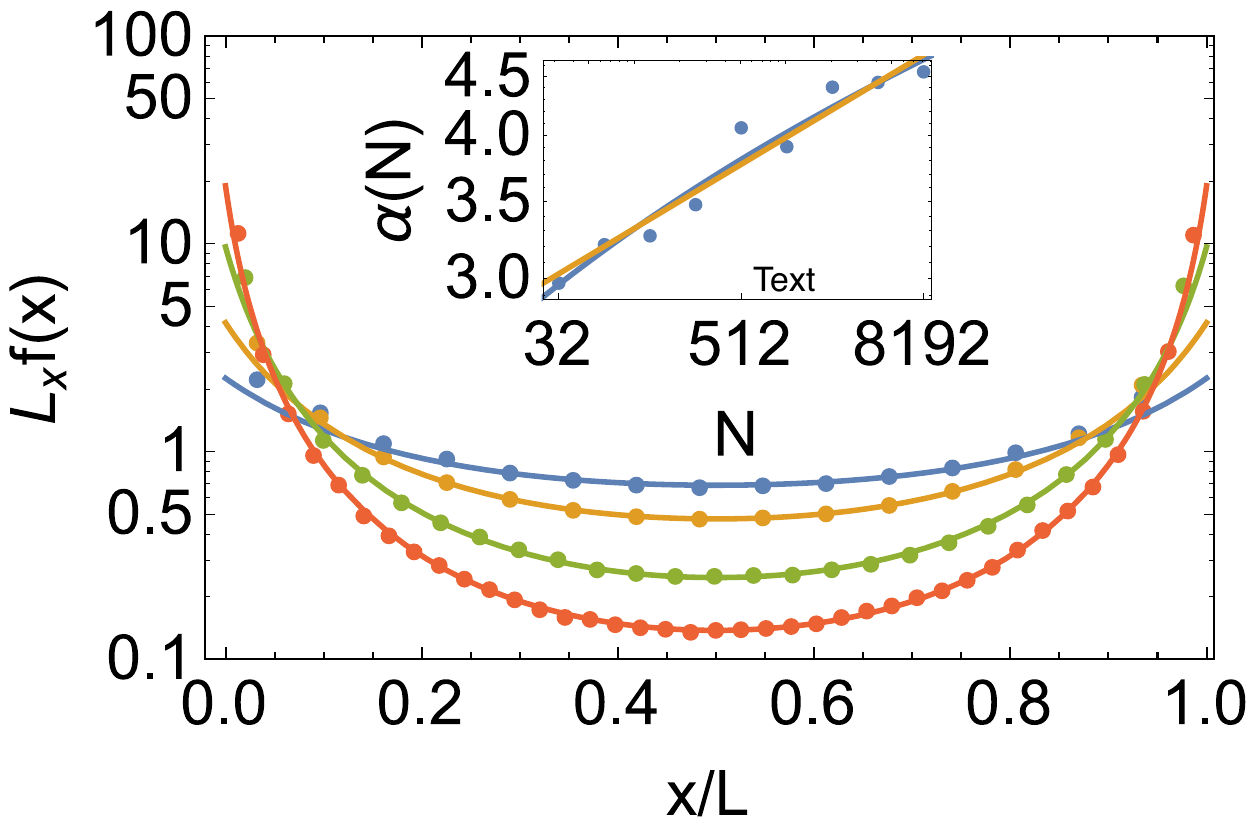}}
\caption{\label{fig:zmstructfit} Numerically obtained
overlap function defined in Eq. (3) of the main text for jammed
two-dimensional unit cells with $n=64$, 256,1024, 8192 (points,
top to bottom), together with the fit based on Eq.
(\ref{eq:supoverlap}) (lines). Inset: Best-fit value of
$\alpha$ as a function of system size, together with
logarithmic and power-law fits to the data. Here the power-law
exponent is $<0.1$.}
\end{figure}

The overlap function, averaged over many realizations of the
disorder, is directly related to $P(\tilde{\kappa})$. Assuming
that the inverse penetration depth of each exponentially
localized zero mode is independently drawn from the
$P(\tilde{\kappa})$, a straightforward calculation detailed later in this section connects $P(\tilde{\kappa})$ to $f(x)$.
Intuitively, the fact that we do not observe strong shifts in
$P(\tilde{\kappa})$ towards more slowly decaying modes with
increasing system size suggests that $H_1(n)$ is a
monotonically decreasing function, and our analytical arguments
presented below confirm that $H_1(n)$ is not bounded from below
by a positive constant. As we show in
Fig.~\ref{fig:zmstructfit}, this expectation is also
numerically confirmed by further simulations of JSP Maxwell
lattices from which free surfaces are cut. This undermines part
of the variational argument \cite{WyartWit2005b,Wyart2005a}
for the jamming transition, which directly connects the lower
bound on $H_1$ to an upper bound on the energetic cost of
potential low-frequency vibrations. We note that this
finding is not restricted to two-dimensional JSP lattices;
directly studying the overlap function for three-dimensional
jammed Maxwell lattices also shows that  $H_1(n)$ is a
monotonically decreasing function of system size.

Without this upper bound, our understanding of the
jamming boson peak is once again uncertain, although we note
that effective-medium calculations (e.g.
Ref. \citenum{DiGiuli2014}) and a very recent variational
argument\cite{Yan2016} also treat the behavior of the density
of states near the jamming transition. Rather than performing a
variational calculation on the effect of introducing free
surfaces, the most recent variational work considers the
dipolar response generated by elongating bonds added to an
isostatic network. The argument, which is supported by
numerical results, assumes that the characteristic volume of
the dipolar response field scales as $1/\Delta Z$, where
$\Delta Z$ is the distance of the lattice from the jamming
transition, and that the magnitude of the response is of
similar magnitude everywhere in this volume and does not decay
strongly with distance. That the response field has a volume
that scales as $1/\Delta Z$ is supported both by direct studies
of the response to bond elongation\cite{Lerner2013sm} and also
recent work looking at the SSS associated with the existence of
a given bond in a hyperstatic packing\cite{Sussman2016}. The
evidence that the magnitude of the dipolar response is roughly
constant in this volume may need further study, as there is
evidence that typical states of self stress have a (stretched-)
exponentially decaying spatial profile in the volume of
interest. The relationship between this observation and
averages over typical dipolar response fields generated by bond
elongation is a subject of current investigation.

We now show how the overlap function can be related to the
distribution of penetration depths obtained earlier. While this may be surprising, given that the topological characterization above dealt with the surface modes of semi-infinite systems and not with the fully liberated systems considered by the cutting argument, we note that the exponentially decaying zero modes of a semi-infinite system are also valid zero modes for a system truncated by a second free surface. Thus, we fully expect the behavior of the overlap function for a single unit cell liberated by parallel free surfaces to be determined by the statistics of the surface modes deduced in the semi-infinite case considered above.

Restricting ourselves to the two-dimensional case, we
approximate the number of zero modes in the cut system by the
surface area of the cut, (since this is proportional to the
number of bonds cut), $n_0\sim L$, and note that the number of
sites in the interval $x_i\in [x,x+dx]$ is $\rho L dx$ where
$\rho$ is the number density of sites in the unit cell.
Finally, we note that the modes $| \brm{u}_{i,\beta} |^2$ are
normalized:
\begin{equation}
\int_0^L |u(x)|^2 \rho L dx = 1.
\end{equation}
If the average magnitude of mode $\beta$ at $x$,
\begin{equation}
u_\beta(x)=(\rho L dx)^{-1}\sum_{x_i\in [x,x+dx]}
\brm{u}_{i,\beta},
\end{equation}
is independent of $x$ (as is the case for the translational
modes) then this normalization sets
\begin{equation}\label{eq:suptrans}
|u_{\text{trans}} (x)|^2= 1/\rho L^2.
\end{equation}
On the other hand, if $u_\beta(x)$ has an exponentially
decaying profile with inverse penetration depth
$\tilde{\kappa}$ then the normalization condition sets
\begin{equation}
|u_{\text{exp}}(x)|^2 = \frac{2 \tilde{\kappa} \left| \exp
\left( -x \tilde{\kappa} \right) \right|^2}{\rho L \left(
1-\exp \left( -2 L \tilde{\kappa} \right) \right)}.
\end{equation}

Of the $L$ zero modes there will be two translational zero
modes, and $(L-2)$ exponentially decaying modes with inverse
penetration depth drawn from the distribution
$P(\tilde{\kappa})$. Averaging over many realizations of the
cut Maxwell-lattice jammed configurations, we have
\begin{equation}\label{eq:supoverlap}
\langle f(x) \rangle \approx\frac{2}{L} |u_{\text{trans}}
(x)|^2 \rho L + \frac{L-2}{L} \langle |u_{\text{exp}} (x)|^2
\rangle \rho L ,
\end{equation}
where
\begin{equation}
\langle |u_{\text{exp}} (x)|^2 \rangle = \int_{-\infty}^\infty
P(\tilde{\kappa}) \frac{2 \tilde{\kappa} \left| \exp \left( -x
\tilde{\kappa} \right) \right|^2}{\rho L \left( 1-\exp \left(
-2 L \tilde{\kappa} \right) \right)}d\tilde{\kappa}.
\end{equation}
Taking the distribution of $\tilde{\kappa}$ to be exponential,
$P(\tilde{\kappa}) \approx (\alpha/2)\exp(-|\alpha
\tilde{\kappa}|)$, the above integral can be written in terms
of the first derivative of the digamma function, $\psi '(z)$:
\begin{equation}\label{eq:supexp}
\langle |u_{\text{exp}} (x)|^2 \rangle= \frac{\psi '\left(
\frac{1+2\alpha x}{2 \alpha L} \right)+\psi '\left(
\frac{1-2\alpha x}{2 \alpha L} \right)}{4 \alpha \rho L^3}
\end{equation}
Substituting Eqs. (\ref{eq:suptrans}, \ref{eq:supexp}) in Eq.
(\ref{eq:supoverlap}) then gives an expression for the overlap
function averaged over many realizations of the disorder. We
plot this in Fig.~\ref{fig:zmstructfit} for different values of
$n$. The inset to Fig.~\ref{fig:khist} shows that the decay
rate of $P(\tilde{\kappa})$, i.e., the value of $\alpha$,
increases with increasing $n$.  Adjusting $\alpha$ to provide
the best fit to the overlaps functions for different $n$ yields
the solid curves in Fig.~\ref{fig:zmstructfit}.  The exponent
$\alpha$ increases from approximately $3$ to $4.5$ as $n$
increases over two orders of magnitude from $32$ to $8192$.
This suggests that, for the two-dimensional JSP, the mean of
$P(\tilde{\kappa})$ may be logarithmically increasing with
system size. Understanding precisely what controls the observed
$\alpha$ for these disordered systems is an open challenge.

\section{Summary\label{sec:disc}}
In summary, we have studied topologically protected boundary
modes and $\kv$-localized Weyl modes in large-unit-cell
lattices derived from model jammed systems. In addition to
computing winding numbers to identify the topological classes
of our lattices, we have extended the formalism in Ref.
\cite{LubenskySun2015} to compute complete sets of exponential
decay profiles for all elements in the nullspace of $\brm{C}$
for large systems with free surfaces. We discovered that
randomized Penrose tilings and jammed unit cells are a rich
source of lattices that can take on a topologically non-trivial
character. Furthermore, the structure of these topologically
modes indicate an interesting inconsistency in an argument
explaining one of the most prominent features of the jamming
transition, pointing towards the need for a more complete
theory. Finally, the close correspondence between the GPT and
JSP, previously documented for their elastic properties
\cite{Stenull2014} and here observed in their topological
characterization, further corroborates the idea that generic
Penrose tilings are useful model systems for jammed matter.

\begin{acknowledgments}
We thank Bryan Chen for making the RigidityPackage available
\cite{RigidityPackage}, with which Figs. \ref{fig:topoglue},
\ref{fig:topoglue2} were made. This work was supported by NSF
under grants DMR-1104707 and DMR-1120901 (TCL and OS) and by
the Advanced Materials Fellowship of the American Philosophical
Society (DMS). TCL is grateful for support from a Simons Fellows grant.
\end{acknowledgments}

\bibliography{RPP1}

\begin{thebibliography}{62}
\bibitem [{foo()}]{footnote2}%
  \BibitemOpen
  \href@noop {} {}\bibinfo {note} {The term isostatic is often used to describe
  lattices with $N_B=dN$. Isostatic lattices are, however, strictly speaking
  Maxwell lattices with additional constraints. See Sec. 2.2.}\BibitemShut
  {Stop}%
\end{thebibliography}


\end{document}